\documentclass[aps,prd,preprint,superscriptaddress,tightenlines,nofootinbib]{revtex4}

\usepackage{graphicx}
\usepackage{dcolumn}
\usepackage{bm}
\newcommand{\invfb}{\mbox{\rm fb$^{-1}$}}
\newcommand{\xp}{\mbox{$x_p$}}
\newcommand{\dedx}{\mbox{$dE/dx$}}
\newcommand{\sigbr}{\mbox{$\sigma\cdot {\cal B}$}}
\newcommand{\br}{\mbox{$\cal B$}}

\newcommand{\cleoiii}{CLEO\,III}
\newcommand{\casmi}{\mbox{$\Xi^-$}}
\newcommand{\cascz}{\mbox{$\Xi^0_c$}}
\newcommand{\lz}{\mbox{$\Lambda$}}
\newcommand{\lc}{\mbox{$\Lambda_c$}}
\newcommand{\kmi}{\mbox{$K^-$}}

\newcommand{\pim}{\mbox{$\pi^-$}}
\newcommand{\pip}{\mbox{$\pi^+$}}
\newcommand{\p}{\mbox{$p$}}
\newcommand{\decays}{\mbox{$\rightarrow$}}
\newcommand{\kstrz}{\mbox{$\overline{K^*}(892)^0$}}
\newcommand{\mevcsq}{\mbox{\rm MeV/$c^2$}}
\newcommand{\mevc}{\mbox{\rm MeV/$c$}}

\newcommand{\gevc}{\mbox{\rm GeV/$c$}}
\newcommand{\ksdec}{\mbox{$\kstrz \to K^-\pi^+$}}
\def\etal{{\em et al.}}

\begin{document}

\preprint{CLNS 03/1834}       
\preprint{CLEO 03-11}         

\title{
Measurement of the decay rate of \cascz\decays\p\kmi\kmi\pip\ relative to \cascz\decays\casmi\pip}



\author{I.~Danko}
\affiliation{Rensselaer Polytechnic Institute, Troy, New York 12180}
\author{D.~Cronin-Hennessy}
\author{C.~S.~Park}
\author{W.~Park}
\author{J.~B.~Thayer}
\author{E.~H.~Thorndike}
\affiliation{University of Rochester, Rochester, New York 14627}
\author{T.~E.~Coan}
\author{Y.~S.~Gao}
\author{F.~Liu}
\author{R.~Stroynowski}
\affiliation{Southern Methodist University, Dallas, Texas 75275}
% ...
\collaboration{CLEO Collaboration} 
\noaffiliation



\date{\today}

\begin{abstract} 
Using the \cleoiii\ detector at CESR, we have measured the branching ratio of
the decay \cascz\ $\to$ \p\kmi\kmi\pip\ relative to \cascz\ $\to$ \casmi\pip.
We find ${\cal B}(\cascz\to\p\kmi\kmi\pip)/{\cal B}(\cascz\to\casmi\pip)$
= 0.35$\pm$0.06(stat)$\pm$0.03(syst).
In the resonant substructure of this mode, we find evidence for 
$\Xi_c^0$ decays to $p\kstrz K^-$, and measure 
${\cal B}(\Xi_c^0\to p \kstrz K^-)\cdot{\cal B}(\kstrz \to K^-\pi^+)
/{\cal B}(\Xi_c^0\to\Xi^-\pi^+)=0.14\pm$0.03(stat)$\pm$0.01(syst) 
and  ${\cal B}(\cascz\ \to \p\kmi\kmi\pip)/{\cal B}(\cascz\ \to \casmi\pip)$=
0.21$\pm$0.04(stat)$\pm$0.02(syst) for the non-\kstrz\ \cascz\ \decays\
\p\kmi\kmi\pip\ decays. This note has the revised numbers on the
branching ratios with improved secondary vertex finding algorithm.

\end{abstract}

\maketitle
In the past decade, singly-charmed baryons 
(consisting of one heavy quark and two light quarks $Qq_1q_2$) 
have been of interest to many phenomenologists working in the 
realm of Heavy Quark Effective Theory~\cite{hqet}. 
The heavy charm quark acts as a heavy nucleus and 
the light di-quark moves around it, analogous to the hydrogen atom. 
The CLEO experiment has 
discovered many new charmed baryons and measured their properties; in particular
it has measured many of their relative branching fractions. 
The study of charmed baryon decays is complicated because they can 
proceed by three distinctly different processes; external $W$-emission, internal 
$W$-decay, and $W$-exchange. Disentangling the contributions of each
of these processes requires the measurement
of as many different decay modes as possible. 

Improvements in the particle identification in the \cleoiii~\cite{cleoiii} 
detector with the 
introduction of the RICH (Ring Imaging CHerenkov) sub-detector~\cite{rich}, 
have made it possible to search
for decay modes previously contaminated by huge combinatorial background. 
This paper concentrates on 
one such decay mode of the \cascz \  (the $csd$ charmed baryon, 
discovered by CLEO~\cite{xic}), namely
\cascz\decays\p\kmi\kmi\pip\ , and its substructure.  
To measure the relative branching fractions
we use \cascz\decays\casmi\pip\ as the normalizing mode.
Charged conjugation is implied throughout the text.
The only previous observation of the $pK^-K^-\pi^+$ final state
was made 
in 1990 by the ACCMOR Collaboration~\cite{accm}, who observed four
\cascz\decays\p\kmi\kstrz\ decays; but there was no information on
their rate.

The data for this analysis were collected 
using \cleoiii\ detector based at the 
Cornell Electron Storage Ring (CESR) taken at and near the 
$\Upsilon(2S)$, $\Upsilon(3S)$, and $\Upsilon(4S)$ resonances.
The integrated luminosity corresponds to 7.2 \invfb .
In the \cleoiii\ detector configuration, the innermost tracking 
device is a four-layer double-sided
silicon vertex detector surrounding the beam pipe. 
Beyond this vertex detector
is the main cylindrical drift chamber~\cite{dr3},
with the inner 16 layers being axial and the outer 31 layers 
having a small stereo angle. The tracking system is 
immersed in a 1.5\,T solenoidal magnetic field and 
measures the momentum and specific ionization (\dedx) of charged particles. 
Outside the drift chamber is the RICH sub-detector~\cite{cleoiii}
consisting of two concentric cylinders.
The inner cylinder comprises LiF-crystal radiators on a 
carbon-fiber shell, while the outer one consists of thin 
multi-wire proportional chambers filled
with a gas mixture of methane and TEA.
Ultra-violet Cherenkov photons interact with the TEA emitting electrons that are
multiplied by the chamber and whose image charge is sensed on pad
detecting planes, thus localizing the photon positions.
Between these two layers is a 16-cm thick gap which allows the 
Cherenkov cone to expand. Surrounding the RICH is the 7800 CsI 
crystal calorimeter for the identification of photons and electrons. 
Beyond the crystal calorimeter is the  superconducting solenoid 
magnet and a muon detector system. The crystal 
calorimeter and the muon detectors are 
not used in this analysis.
  
To identify hadrons, we combine information on the specific ionization
(\dedx) measured in the drift chamber and 
likelihoods obtained from the RICH detector.
The RICH likelihood is formed for each hypothesis ($i = \pi$, $K$, \p ) 
by using the measured positions of photons that are located
within 3 standard deviations of their expected positions for the measured 
track momentum.
The likelihood function for each particle hypothesis, {\it i}, is defined as,
\begin{equation}
L_i = \prod_{j=1}^{N^{obs}}[G(\theta^{obs}_j|\theta^{exp}_i) + B],
\end{equation}
where $G$ is a Gaussian-like probability function
of observing the $j^{th}$ photon at an angle $\theta^{obs}_j$ with
respect to the expected  
Cherenkov
angle $\theta^{exp}_i$ for particle type {\it i},
and $B$ is a flat background probability function. 
The details of RICH identification 
are given elsewhere~\cite{blusk}.
Similarly, using the \dedx\ information, we 
construct a quantity for 
the different hypotheses ($i = \pi$, $K$, \p ) as
$S_{i}$, which is the difference between the measured and expected \dedx\
for that hypothesis, expressed in units of its standard deviation.
The RICH and \dedx\ information for each pair of hypotheses is then
combined to form $\chi^2$ functions, for example,
\begin{equation}
\Delta\chi^2 (p - \pi)  = 
-2{\rm log}L_p +2{\rm log}L_{\pi}  +
S_{p}^2 - S_{\pi}^2 .
\end{equation}
In this example real protons peak at negative values
of $\Delta\chi^2 (p - \pi)$ whereas real pions tend to have 
positive values.
If there is no RICH information available, or if the 
particle's momentum is less than 1.0 \gevc\ (for a proton), 
or 0.5 \gevc\ (for a kaon),
we use only \dedx\ information to form the $\Delta\chi^2$ function.

All the primary charged tracks 
are required to have a distance of closest approach to the beam position
in the $\hat{r}-\hat{\phi}$ plane of less than 5\,mm 
and of less than 5\,cm along the $\hat{z}$ axis. 
Typical beam dimension is ~300 microns, 100 microns, and 10mm in x,y,
and z, respectively. 
In the CLEO environment charmed baryons do not have 
well separated decay vertices.
We require that the scaled momentum of the charmed baryon
candidate, $x_p$, be greater than 0.5. Here 
$\xp = P/\sqrt{E_b^2 - M^2}$,
{\it P} and {\it M} are the momentum and mass of the
candidate, and $E_b$ is the beam energy.
This requirement dramatically suppresses 
combinatorial background, and dictates that the observed charmed baryons
are the result of continuum production rather than from $B$ meson decays.

\begin{figure}
\includegraphics*[width=3.75in]{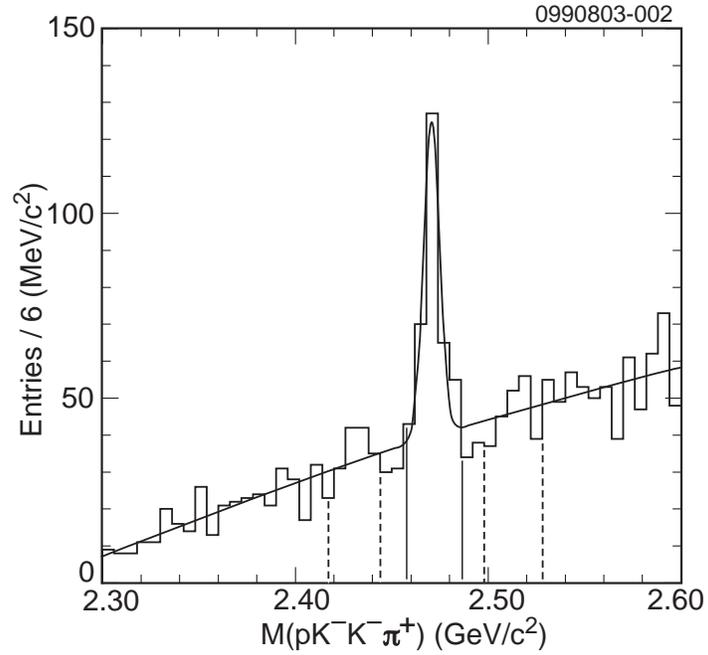}
\caption{ Invariant mass distribution of \p\kmi\kmi\pip\ candidates in
\cleoiii\ data. The fit to the above mass distribution yields 148$\pm$18 signal events.
The signal band (2458.3 $-$ 2483.1 \mevcsq ) is defined within the solid lines and the 
low (2417.6 $-$ 2442.4 \mevcsq) and high ( 2498.3 $-$ 2523.1 \mevcsq) side bands 
are defined by the dashed lines.}
\label{fig:pkkpi}
\end{figure}

Candidates for the \cascz\decays\p\kmi\kmi\pip\ decay are reconstructed by 
combining one proton candidate, two kaon candidates, and a pion candidate.
For the proton identification, we require $\Delta \chi^2 (p-K)<-4$ and
$\Delta \chi^2 (p - \pi)<-4$. 
Similarly for each kaon identification, we require
$\Delta \chi^2 (K-p)<-4$ and $\Delta \chi^2 (K - \pi)<-4$. 
Pions are selected with a
loose \dedx\ criteria $|S_{\pi}| \,<$ 5.
All charged tracks are required to have momenta in excess of 100 \mevc. 
Once the four charged tracks are selected we kinematically constrain them to come from a 
common vertex.
The invariant mass distribution of the \p\kmi\kmi\pip\ candidates is shown in
Fig.~\ref{fig:pkkpi}. The data are fit to a Gaussian signal function 
and a second-order polynomial background shape. The fit yields
a signal of $148\pm18$ events at a mass consistent with previous measurements of the
$\Xi_c^0$~\cite{pdg} and a fitted width of 4.1$\pm$0.5 \mevcsq, consistent with 
expected resolution of 4.5 \mevcsq\ obtained using a
GEANT-based Monte Carlo simulation~\cite{geant}. 

\begin{figure}
\includegraphics*[width=3.75in]{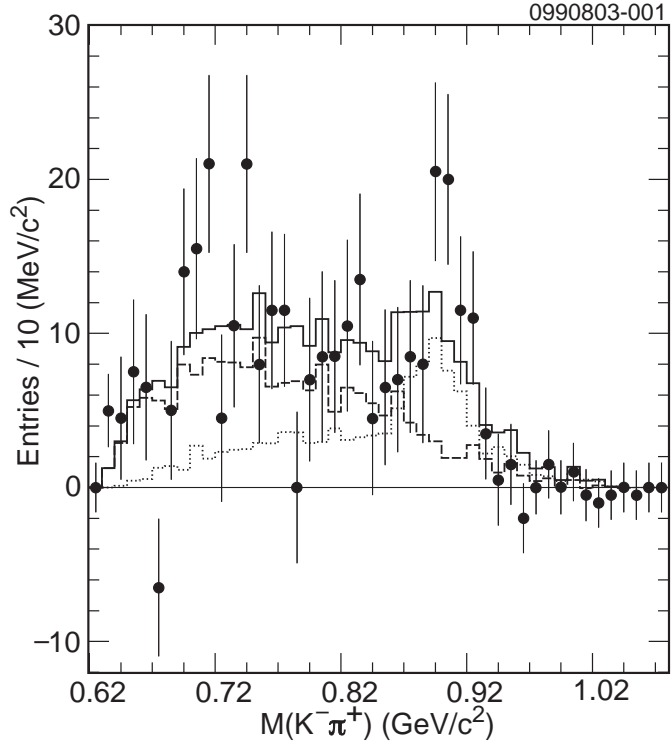}
\caption{The \kmi\pip\ invariant mass in \cleoiii\ data for \cascz\
candidates within 3 standard deviations of the PDG value. The two sideband 
contributions have been subtracted. Dots with error bars are data points,
the dashed histogram is the non-\kstrz\ contribution and the dotted
histogram is the resonant contribution. The solid line histogram
is the sum of the two contributions.
}
\label{fig:kstr}
\end{figure}

In this multi-body final state, \p\kmi\kmi\pip, 
we also search for resonant sub-structure
\kstrz\ by calculating the invariant mass of each of the kaon candidates
combined with the pion. Figure~\ref{fig:kstr} shows the sideband-subtracted
$K^-\pi^+$ invariant mass using those $p K^- K^-\pi^+$ 
combinations within 3 standard deviations of the
$\Xi_c^0$ mass peak. Candidates for the signal (sidebands) are 
selected within the mass region of 2458.3 $-$ 2483.1 \mevcsq\ 
(2417.6 $-$ 2442.4 or 2498.3 $-$ 2523.1 \mevcsq), as shown in
Fig.~\ref{fig:pkkpi}. The low mass tail in this distribution is due to
the fact that we combine both kaons with a pion. The correct combination
appears in the peak region, and the incorrect once forms the low mass tail.
This distribution is fit to the sum of two shapes. The first one is
the \kstrz\ signal shape, which is generated using a 3-body phase
space simulation of \cascz\decays\p\kmi\kstrz, with \kstrz\decays\kmi\pip.
For the second , we use a four-body non-resonant simulation of \cascz\decays\p\kmi\kmi\pip,
since the data show no evidence of any other narrow resonances. In particular,
we also searched for the two-body decay of the \cascz\decays\lz(1520)\kstrz,
where \lz(1520)\decays\p\kmi\ and \kstrz\decays\kmi\pip, and found no evidence
 for this mode. Moreover, broad
resonances do not yield a statistically significant difference in shape than
that from non-resonant production. We therefore model all the non-\kstrz\
contributions using this non-resonant production model.
 We fit this plot to the sum of two shapes from
Monte Carlo simulation, one obtained for non-\kstrz\ 
\cascz\decays\p\kmi\kmi\pip\ decays (dashed histogram) , and the other for resonant 
\cascz\decays\p\kmi\kstrz\ decays (dotted histogram), where \kstrz\decays\kmi\pip.
In the fit the two normalizations are constrained to add to unity. 
The statistics are too poor to extract any possible interference effects in the invariant 
mass distribution, and is therefore not considered here.
The measured resonant and non-\kstrz\ fractions are found to be 0.39$\pm$0.06(stat)
and 0.61$\pm$0.06(stat), respectively.  
Thus, of the total 148$\pm$18 fitted \cascz\ (\decays\p\kmi\kmi\pip) 
candidates 58$\pm$11(stat) events are contributed by resonant \p\kmi\kstrz\
decays and 90$\pm$14(stat) events are contributed by non-\kstrz\ decays. 
The $\chi^2$/dof of the fit (shown in Fig.~\ref{fig:kstr}) to
the resonant and non-\kstrz\ components is found to be 44/45, 
indicating that our fitting function, which includes a contribution from only
one resonance, is a satisfactory one within the available statistics,
and we believe that the systematic uncertainty may be neglected.

The reconstruction efficiency of these two final states are estimated
using the Monte Carlo simulation. Within the kinematic region $x_p > 0.5$, 
we find that both final states have a reconstruction efficiency of 
(23$\pm$1) $\%$. The corresponding efficiency-corrected
measured cross-sections times branching fractions (\sigbr) are
91$\pm$12(stat)$\pm$8(syst), 37$\pm$7(stat)$\pm$3(syst), 
and 54$\pm$9(stat)$\pm$5(syst) fb for all \p\kmi\kmi\pip, 
\kstrz\ resonant (\kstrz\decays\kmi\pip\ only) 
and non-\kstrz\ \p\kmi\kmi\pip, respectively, where all the measurements refer
to that part of the momentum spectrum with $x_p > 0.5$.

\begin{figure}
\includegraphics*[width=3.75in]{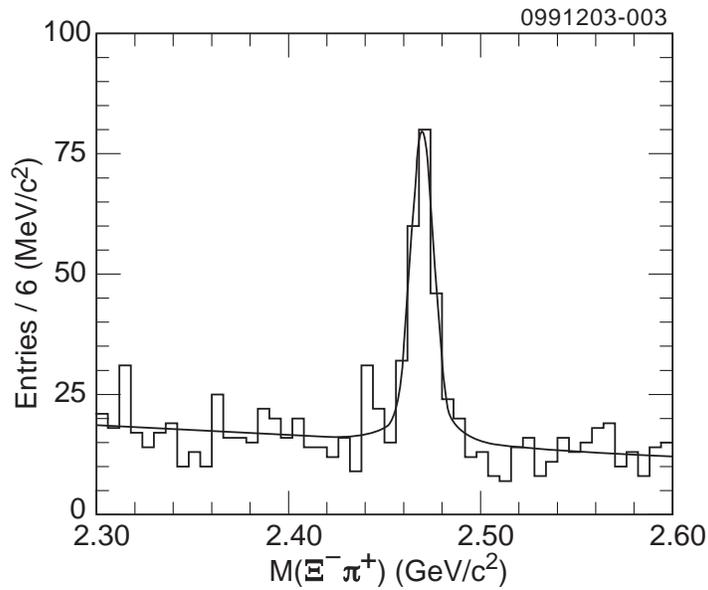}
\caption{Invariant mass distribution of \casmi\pip\ candidates in
\cleoiii\ data. The fit to the mass distribution yields 182$\pm$18
signal events and the fitted mean is consistent with the nominal 
\cascz\ PDG mass~\cite{pdg}.}
\label{fig:ximpi}
\end{figure}

As the production cross-section of $\Xi_c^0$ baryons is unknown, we
do not have a measure of the absolute branching fraction of any $\Xi_c^0$ mode.
Instead, we measure the branching ratios of these new modes with
respect to that of the well-established decay $\Xi_c^0 \to \Xi^-\pi^+$, where \casmi\decays\lz\pim. Both the \casmi\ and its daughter \lz\ have 
long flight paths, with $c\tau$ values of 4.91 and 7.89\,cm, respectively. 
Therefore, in the \cascz\decays\casmi\pip\ decay chain we have two vertices
significantly detached from the beamspot. 
The \lz\ sample is selected by vertexing two oppositely charged tracks.
The protons from the \lz\ decays, which have the higher momentum of the two 
daughters, are required to be consistent 
with a proton hypothesis ($\Delta \chi^2 (p-K)$ and $\Delta \chi^2 (p-\pi)$ $<$ 0).
Background is further rejected by requiring the daughter tracks from
the \lz\ to be inconsistent with coming from the beam interaction point.
The \lz\ candidates within 5 \mevcsq\ (3 standard deviations) of the nominal PDG~\cite{pdg} 
mass (1115.68 \mevcsq) are then kinematically constrained to this mass and combined 
with an appropriately charged track to form the \casmi\ candidate. The \lz\ decay vertex 
is required to be at a greater distance from the beamspot than the \casmi\ decay vertex. 
The \casmi\ candidates are also required to have a flight distance of 3\,{ mm} or
more before decaying. Pions from the \casmi\ baryons are
required not to come from the interaction point, by requiring the
$\chi^2$ of the fit (if forced to come from the interaction point) to be greater than 3. 
Those \casmi\ candidates within 7 \mevcsq\ (3 standard deviations) of the
PDG~\cite{pdg} mass (1321.31 \mevcsq) are kinematically
constrained to this mass and are used for further analysis. 
Finally a charged track consistent with the pion hypothesis is combined
with the \casmi\ candidate to reconstruct the \cascz\ candidate.
A fit to the $\Xi^-\pi^+$ invariant mass distribution returns a signal yield 
of $182\pm18$ and a mass consistent with previous measurements as
shown in Fig.~\ref{fig:ximpi}. The \cascz\decays\casmi\pip\ reconstruction efficiency 
is measured to be (9.8$\pm$0.3) $\%$. The measured branching fraction times 
the cross section (\sigbr) for the
\casmi\pip\ mode is 260$\pm$26(stat)$\pm$23(syst) fb for $x_p > 0.5$. 

The measured relative branching fractions of the \cascz\decays\p\kmi\kmi\pip\ modes 
are tabulated in Table~\ref{table:ratio}.

\begin{small}
\begin{table}[h]
\begin{center}
\caption{Measured branching fractions of the
\cascz\decays\p\kmi\kmi\pip\ mode relative to that for \cascz\decays\casmi\pip. 
The errors after the values give the statistical and systematic uncertainties, respectively.}
\label{table:ratio}
\begin{tabular}{l c} \hline
Ratio of modes & Relative branching fraction   \\ \hline
$\frac{\br(\cascz\decays\p\kmi\kmi\pip)}{\br(\cascz\decays\casmi\pip)}$ & 0.35 $\pm$ 0.06 $\pm$0.03 \\
$\frac{\br(\cascz\decays\p\kmi\kstrz)\cdot\br(\ksdec)}{\br(\cascz\decays\casmi\pip)}$ & 0.14 $\pm$ 0.03$\pm$0.01 \\ 
$\frac{\br(\cascz\decays\p\kmi\kmi\pip)~{\rm No~\kstrz}}{\br(\cascz\decays\casmi\pip)}$&0.21 $\pm$0.04 $\pm$0.02\\ \hline
\end{tabular}
\end{center}
\end{table}
\end{small}

We investigated several sources of uncertainty, including background
shapes, signal width, Monte Carlo statistics, charged particle
identification, and \lz\ and \casmi\ reconstruction. 
The dominant uncertainties arise from the \casmi\pip\ mode, due to the
reconstruction of the displaced vertices.

	To estimate the uncertainty due to our assumptions in the
shape of the background, we tried both first and second order
polynomials to describe its shape. The fitted yield changed by 3$\%$,
which we take as our systematic uncertainty from this source. 
The systematic uncertainty due to the imperfect understanding 
of the signal resolution is taken as the difference in \cascz\ signal 
yield using a floating width and a width fixed to the value found 
from simulation (3$\%$ for the branching ratios, 6$\%$ and 8$\%$ 
for the \p\kmi\kmi\pip\ and the \casmi\pip\ modes, respectively).
We assign 4$\%$ and 5$\%$ systematic uncertainties to the
\cascz\decays\p\kmi\kmi\pip\ and \cascz\decays\casmi\pip\ decay modes,
respectively, due to the finite statistics of the Monte Carlo samples.
As the number of tracks for the observed (numerator) and the 
normalizing (denominator) modes are the same, the 1\%
per track uncertainty in  basic track-finding cancels in the calculation of the 
branching ratio, as does the uncertainty in the luminosity (2\%).
Systematic uncertainties in the charged particle identification are 
investigated using samples of protons and kaons from the
\lz\decays\p\pim\ and \lc\decays\p\kmi\pip\ modes, respectively. 
The study demonstrated that the uncertainties in the particle 
identification for protons and kaons in the relevant momentum range are 3.2$\%$
and 2.4$\%$, respectively. As there are two kaons in the final state, 
we assign 4.8$\%$ uncertainty to the kaon identification.
Based on a study of displaced vertex finding in data and Monte Carlo 
we assign systematic uncertainties of 6$\%$
to the \lz\  finding and 2.8$\%$ to the \casmi\ finding; these numbers
include the extra uncertainty in track-finding for low momentum, large
impact parameter tracks. All systematic uncertainties are summarized in 
Table~\ref{table:syst}. The total systematic uncertainty, obtained by
adding the individual contributions in quadrature, is
9$\%$, and considerably less than the statistical uncertainty.

\begin{small}
\begin{table}[h]
\begin{center}
\caption{ Systematic Uncertainties: The systematic errors listed under
Ratio are for the relative branching fractions and the errors
tabulated in the third and fourth column are \p\kmi\kmi\pip\ and 
\casmi\pip\ modes, respectively. }
\label{table:syst}
\begin{tabular}{ l |c |c |c} \hline
Source                             & \multicolumn{3}{|c}{Uncertainty ($\%$)}   \\ \hline
                                   & Ratio      &\p\kmi\kmi\pip & \casmi\pip \\ \hline
Luminosity                       &  -         &    2        & 2 \\ 
Track Reconstruction		   &  -         &    4        & 4 \\                   
Background shape                   & 3          &    3        & 3 \\ 
Signal Width (Monte Carlo)         & 3          &    6        & 8  \\
\p\kmi\kmi\pip\ (Monte Carlo statistics)& 4	&    4	      & - \\
\casmi\pip\ (Monte Carlo statistics)   & 5	&    - 	      & 5  \\
Proton ID                          & 3.2        &    3.2      & 3.2 \\
Kaon  ID (two kaons)               & 4.8        &    4.8      & - \\
\lz\ reconstruction                & 6          &    -        & 6 \\ 
\casmi\ reconstruction             & 2.8        &    -        & 2.8 \\ \hline
Total                              & 9.8        &    9        & 9    \\ \hline
\end{tabular}
\end{center}
\end{table}
\end{small}

 In conclusion, we have measured the branching fraction
for \cascz\decays\p\kmi\kmi\pip\ relative 
to \cascz\decays\casmi\pip. We find that 39$\pm$6(stat)$\%$ of the 
signal proceeds via the
resonance sub-structure \p\kmi\kstrz, where \kstrz\decays\kmi\pip\ 
with the remainder being
the non-\kstrz\ decay \p\kmi\kmi\pip.
Results are given as a ratio normalized to the \casmi\pip\ rate. 
The measured branching ratios for 
$\frac{\br(\p\kmi\kmi\pip)}{\br(\casmi\pip)}$, 
$\frac{\br(\p\kmi\kstrz)\cdot\br(\ksdec)}{\br(\casmi\pip)}$,
and $\frac{\br(\p\kmi\kmi\pip)~{\rm No~\kstrz}}{\br(\casmi\pip)}$ 
are 0.35$\pm$0.06(stat)$\pm$0.03(syst), 0.14$\pm$0.03(stat)$\pm$0.01(syst), and 
0.21$\pm$0.04(stat)$\pm$0.02(syst), respectively.
This is the first measurement of a $\Xi_c^0$ decay mode where both of 
the $s$ quarks in the final state are part of mesons. It is possible
that such four-body decays proceed via external $W$-decay, internal $W$-decay
or $W$-exchange decay diagrams. Resonant decays such as \p\kstrz\kmi,
which have no  $\pi^+$ in the final state, cannot decay via external $W$-decay. 
Their observation is not surprising, as many such
$\Lambda_c^+$ and $\Xi_c^0$ decays modes have been discovered, 
and is a further indication that external $W$-decay diagrams do not dominate
in charmed baryon decays. 

We gratefully acknowledge the effort of the CESR staff in providing us with
excellent luminosity and running conditions.
M. Selen thanks the Research Corporation, 
and A.H. Mahmood thanks the Texas Advanced Research Program.
This work was supported by the 
National Science Foundation, 
and the
U.S. Department of Energy.

\end{document}